\documentclass{aa}
\usepackage[varg]{txfonts}
\usepackage{graphicx}
\usepackage{amsmath}
\usepackage{natbib}
\bibpunct{(}{)}{;}{a}{}{,}
\def\ltsima{$\; \buildrel < \over \sim \;$}
\def\lsim{\lower.5ex\hbox{\ltsima}}
\def\gtsima{$\; \buildrel > \over \sim \;$}
\def\gsim{\lower.5ex\hbox{\gtsima}}
\usepackage[utf8]{inputenc}

\begin{document}

\title{X-ray study of high-and-low luminosity modes and peculiar low-soft-and-hard activity in the transitional pulsar XSS J12270$-$4859}
\titlerunning{X-ray study of transitional pulsar XSS J12270$-$4859}

\author{A. Miraval Zanon\inst{1,}\inst{2}, S. Campana\inst{2},
  A. Ridolfi\inst{3,}\inst{4}, P. D'Avanzo\inst{2} \and F. Ambrosino\inst{5,}\inst{6}}

\institute{Universit\`a dell'Insubria, Dipartimento di Scienza e Alta Tecnologia, Via Valleggio 11, I-22100 Como, Italy
  \and INAF, Osservatorio Astronomico di Brera, Via E. Bianchi 46, I-23807 Merate (LC), Italy\\
  e-mail: \texttt{arianna.miraval@inaf.it}
    \and INAF, Osservatorio Astronomico di Cagliari, Via della Scienza 5, I-09047 Selargius (CA), Italy
  \and Max-Planck-Institut f$\ddot{\rm u}$r Radioastronomie, Auf dem H$\ddot{\rm u}$gel 69, D-53121 Bonn, Germany
   \and INAF-IAPS, Via del Fosso del Cavaliere 100 Roma, I-00133, Italy
   \and Sapienza University of Rome, Piazzale Aldo Moro 5 Roma, I-00185, Italy
}
\date{Accepted 2020 January 25}

\abstract{ XSS J12270$-$4859 (henceforth J12270) is the first low-mass X-ray binary to exhibit a transition, taking place at the end of 2012, from an X-ray active state to a radio pulsar state. The X-ray emission based on archival \textit{XMM-Newton} observations is highly variable, showing rapid variations ($\sim$ 10 s) from a high X-ray luminosity mode to a low mode and back. A flaring mode has also been observed.
X-ray pulsations have been detected during the high mode only. In this work we present two possible interpretations for the rapid swings between the high and low modes. In the first scenario, this phenomenon can be explained by a rapid oscillation between a propeller state and a radio-ejection pulsar state, during which the pulsar wind prevents matter from falling onto the neutron star surface. 
In the second scenario, a radio pulsar is always active, the intra-binary shock is located just outside the light cylinder in the high mode, while it expands during the low mode. 
At variance with other transitional pulsars, J12270 shows two instances of the low mode: a low-soft and low-hard mode. Performing an X-ray spectral analysis, we show that the harder component, present in the low-hard spectra, is probably related to the tail of the flare emission. This supports the understanding that the flare mechanism is independent of the high-to-low mode transitions.}
\maketitle 

\section{Introduction}
Millisecond pulsars (MSPs) are believed to form in binary systems, in which the companion star transfers matter and angular momentum onto the neutron star (NS) surface. 
During this phase, the NS is spun-up to very short spin periods (a few ms) and the binary system appears as a low-mass X-ray binary (LMXB). This process, called recycling, can last up to a few Gyr \citep{2013IAUS..291..137T}. During this phase, the accreted mass ($\gsim$ 0.1-0.2 M$_{\odot}$) likely causes the decay of the magnetic field down to 10$^7$-10$^8$ G. When the mass accretion stops, the NS can shine as a radio and $\gamma$-ray pulsar. The MSP recycling scenario (\citealt{1974SvA....18..217B}; \citealt{1975A&A....39...61F}; \citealt{1982Natur.300..728A}) is supported by the detection of few hundreds of radio MSPs in binaries.

 The recycling scenario received strong support from the discovery of coherent X-ray pulsations during type I X-ray bursts first \citep{1996ApJ...469L...9S} and then by the discovery of coherent pulsations during the outburst of (faint) X-ray transients \citep{1998Natur.394..344W}.
The discovery of transitional pulsars (TPs) brought additional support to the recycling model. TPs are binary pulsars swinging between a radio millisecond pulsar state and an X-ray bright state. This X-ray active state does not correspond to a typical outburst from a transient LMXB. Instead, the X-ray luminosity in the X-ray active state is quite dim ($L_{X (0.3-10 \ {\rm keV})} \sim$ 10$^{33}$ erg s$^{-1}$) and stable over several years. Fast and erratic time variability is observed during the X-ray active state, with luminosity changing by a factor of $\sim 5-7$ and transition time scales as short as $\sim 10$ s. X-ray pulsations are detected only during the high mode of this active state (see also below).
Only three confirmed systems are currently known together with four candidates (\citealt{2018arXiv180403422C}; \citealt{2019A&A...622A.211C}). 
IGR J18245$-$2452 (J18245 in the following) was discovered by INTEGRAL/ISGRI \citep{2013ATel.4925....1E} and is located in the globular cluster M28 at a distance of $\sim$ 5.5 kpc \citep{2013ATel.4927....1H}. In 2013, the binary showed a faint LMXB outburst and X-ray pulsations at 3.9 ms were discovered with \textit{XMM-Newton}, classifying it as an accreting millisecond X-ray pulsar \citep{2013Natur.501..517P}. J18245 turned into a millisecond radio pulsar two weeks after the end of the outburst.
PSR J1023+0038 (J1023 in the following) was originally classified as a peculiar magnetic cataclysmic variable \citep{2002PASP..114.1359B} and was the first object known for the transition between an X-ray active state and a radio pulsar state \citep{2009Sci...324.1411A}. In 2007, it was discovered with the Green Bank Telescope as a fast-spinning radio pulsar (hence a NS) with a rotational period of 1.69 ms \citep{2009Sci...324.1411A}. Around June 2013, J1023 was not detected any longer in the radio band, while its $\gamma$-ray flux increased by a factor of $\sim$ 5 \citep{{2014ApJ...790...39S}}. This state has persisted up to the present day (November 2019).
X-ray pulsations were detected both during the radio pulsar state \citep{2010ApJ...722...88A} and in the X-ray active state. Using the fast optical photometer SiFAP at the Telescopio Nazionale Galileo (TNG), J1023 was also recently discovered to be an optical pulsar \citep{2017NatAs...1..854A}. This is the first case of a MSP in which pulsed emission in the optical band has been detected. The optical pulsations were also observed with the fast photon counter Aqueye+, mounted at the Copernico telescope in Asiago \citep{2019MNRAS.485L.109Z}. 

\subsection{XSS J12270$-$4859}
XSS J12270$-$4859 (J12270 henceforth) was discovered as a hard X-ray source during the \textit{Rossi X-ray Timing Explorer} (RXTE) survey \citep{2004A&A...423..469S} but its nature was unclear for a few years.  Initially, the source was thought to be a cataclysmic variable (CV), similarly to J1023 \citep{2006A&A...459...21M, 2008A&A...487..271B}, owing to the presence of double-horned emission lines, typical of an accretion disc. Further observations showed a peculiar X-ray variability with an unusual dipping and flaring behaviour (\citealt{2009PASJ...61L..13S}; \citealt{2010A&A...515A..25D}). Therefore, several authors suggested that J12270 was instead a NS in a LMXB (\citealt{2009MNRAS.395..386P};  \citealt{2009PASJ...61L..13S}; \citealt{2010A&A...515A..25D}; \citealt{2011MNRAS.415..235H}). 
The detection of J12270 by {\it Fermi}-LAT \citep{2010A&A...515A..25D} further garbled the identification of J12270 within the Galactic zoo.

J12270 had a stable emission at all wavelengths from discovery until the end of 2012, when a decrease in its X-ray flux occurred (\citealt{2014MNRAS.441.1825B}; \citealt{2014ApJ...789...40B}). 
A radio pulsar, PSR J1227$-$4853, was discovered in 2014 with the Giant Metrewave Radio Telescope (GMRT) at 607 MHz, and associated with J12270 \citep{2014ATel.5890....1R}, proving the transitional pulsar nature of J12270. \cite{2015ApJ...800L..12R} found that PSR J1227$-$4853 is spinning at 1.69 ms and has a relatively low magnetic field of 1.4 $\times$ 10$^8$ G. $\gamma$-ray pulsations, after the transition, were detected using $\sim$1 year {\it Fermi}-LAT data (\citealt{2015ApJ...806...91J}).

The pulsar signal is eclipsed for a large fraction of the orbit ($\sim$ 40$\%$) at 607 MHz, probably due to free-free absorption by the intra-binary plasma emanated from the companion star.
During the radio pulsar state the X-ray emission is strongly modulated at the $\sim$~6.9~hr orbital period \citep{2015MNRAS.454.2190D}. 
The companion mass is in the range 0.17-0.46 M$_{\odot}$ and the distance, estimated from the dispersion measure, is $\sim 1.4$ kpc. 
An optical polarimetric analysis was performed using the ESO New Technology Telescope (NTT) at La Silla (Chile) during the radio pulsar state. \citet{2016A&A...591A.101B} did not detect any significant polarisation from the system, with 3-$ \sigma $ upper limit of 1.4\% (in R-band), as expected being the source in its millisecond radio pulsar state. 
J12270 is the first binary for which it has been possible to observe the transition from an X-ray active state to a radio pulsar state. 

After the detection of a MSP, (back) searches for pulsations at other wavelengths led to the detection of significant X-ray pulsations in the \textit{XMM-Newton} data (\citealt{2015MNRAS.449L..26P}). 
During the X-ray active state, J12270, like J1023, shows strong X-ray variability, swinging between three different modes: a high luminosity mode in which X-ray pulsations are detected at an rms of (7.7 $\pm$ 0.5)$\%$; a low luminosity mode, during which no pulsations are detected (with an upper limit on rms amplitude of 5.9$\%$); and a flaring mode with no pulsations too \citep{2015MNRAS.449L..26P}. Transitions between high and low modes are rapid, with a timescale of $\sim$ 10 s \citep{2013A&A...550A..89D}. At variance with any other TP, J12270 shows two different low modes: low-soft and low-hard. As presented in \citet{2010A&A...515A..25D, 2013A&A...550A..89D} soft- and hard-low modes stand apart in the intensity-hardness ratio diagram. 

In this paper, two possible physical scenario are presented to interpret the puzzling behaviour exhibited during the X-ray active state: one involving a transition between a propeller (see Section 2.1) and a radio pulsar (\citealt{2016A&A...594A..31C};  \citealt{2014ApJ...795...72L}) and the other one considering an active radio pulsar all the time (\citealt{2014MNRAS.444.1783C}; \citealt{2014ApJ...785..131T}; \citealt{2019arXiv190410433P}; \citealt{Campana2019}).
Using \textit{XMM-Newton} observations carried out during both the X-ray active state and radio pulsar state, and selecting low, high (and flare) X-ray emission modes, we test these scenarios on J12270. In Section 2 we describe the two physical scenarios and their relative spectral models. In addition, we investigate the behaviour between low-soft and low-hard modes through spectral analysis, looking if the low-hard mode spectra can be accounted for by the addition of a further spectral component.
In Sections 3, 4, and 5 we describe our data set and the related spectral analysis. 
In Section 6, we discuss the results and we present future perspectives.

\section{Two possible physical scenarios}
In this section, we present two possible physical scenarios that could explain the rapid variability between high and low luminosity modes.

\subsection{Pulsations due to accretion onto the NS surface in a propeller-ejection scenario}

In this scenario the X-ray active state is characterised by two different accretion modes: propeller and radio-ejection. 
This scenario has been developed for J1023 and has been shown to properly account for its properties (\citealt{2016A&A...594A..31C}).
In the propeller, the inflowing disc matter is halted at the magnetospheric boundary, acting as a centrifugal barrier. Converting the high mode observed luminosity to a mass accretion rate with a typical efficiency of $20\%$, we would obtain a magnetospheric radius of $r_{\rm m}\sim 105$ km (using a factor $\xi$=0.5 for the conversion from spherical to disc accretion \citealt{2018A&A...610A..46C}; see Table \ref{table:1}). This is larger than the light cylinder radius ($r_{\rm lc} = c P/2 \pi$= 80 km, where $P$ is the NS rotational period), indicating that matter should not even reach the magnetosphere and it should instead be expelled from the system.
To force the magnetosphere toward the NS, we must suppose the presence of an advection-dominated flow, in which the dissipated energy is stored rather than being irradiated \citep{1994ApJ...428L..13N, 1996ApJ...462..136N}. 
 This results in a larger disc pressure, allowing matter to reach the magnetospheric boundary. A small fraction of this material might then leak through the centrifugal barrier, reach the NS surface and generate the X-ray pulsations. In this picture, this corresponds to the high mode (\citealt{2016A&A...594A..31C}).
If  the accretion rate decreases, the disc pressure decreases and the magnetospheric radius expands. When this radius becomes comparable to the light cylinder radius, the radio-ejection mode sets in and matter is expelled (e.g. \citealt{1998A&ARv...8..279C};  \citealt{2001ASPC..234..237B}).
This corresponds in our picture to the low mode.
The timescale for this high-low transition (and vice versa) is the free-fall timescale close to the light cylinder, which nicely fit the observed timescale (\citealt{2018A&A...611A..14C}; \citealt{2019arXiv190602519V}).
A radio pulsar, even if reactivated, will be hardly observable due to the large amount of ionised material present in the system.
Finally, the radio pulsar quiescent state is attained when the mass accretion rate further decreases, the pulsar is able to eject all the incoming matter and the system emits in the X-ray band with a flux one order of magnitude lower (see Table \ref{table:1}). 
 We note that this scenario can account for the high and low luminosity modes only. Flares might originate from a mechanism unrelated to the magnetospheric boundary transitions, such as magnetic reconnection in the disc or in the companion (e.g. \citealt{2004ApJ...601..474C}; \citealt{2003ApJ...582..369Z}).


\begin{table*}
\centering
\caption{Estimate of mass loss rate from X-ray luminosity and characteristic radii of J12270. We assume B $\sim$ 1.36 $\times$ 10$^8$ G \citep{2015ApJ...800L..12R}.}              
\label{table:1}      
\begin{tabular}{c c c c}          
\hline\hline                        
Parameter &  High mode & Low mode & Quiescence\\   
\hline 
$L_{X \rm(0.3-10 \ \rm keV)}$ (erg s$^{-1}$)  & 4.2 $\times$ 10$^{33}$ & 6.2 $\times$ 10$^{32}$ &1.1 $\times$ 10$^{32}$ \\
$\dot{M}$ (g s$^{-1}$) & 2.3 $\times$ 10$^{13}$& 3.3 $\times$ 10$^{12}$ & -$^{(*)}$\\
$r_{\rm m}^{(\dagger)}$ (km) &105 &181 &-$^{(*)}$ \\
$r_{\rm c}^{(**)}$ (km) &24 &24 & 24 \\
$r_{\rm lc}$ (km)&80 & 80 & 80\\
State (propeller/ejection model) & propeller & radio-ejection & radio pulsar \\
\hline

\\
\end{tabular}\\
$^{(\dagger)} $ derived as $L_XR/GM$, where $M$ and $R$ are the NS mass and radius, $G$ is the gravitational constant; $^{(*)}$ we do not compute $\dot{M}$ and $r_{\rm m} $ for the quiescent state because the X-ray luminosity is not related to mass accretion;
$^{(**)}$ $r_{\rm c}$ is the corotation radius
\end{table*}

This physical scenario (which is not a detailed model, nor it is unique) can be translated into a spectral model to be tested by fitting X-ray data \citep{2016A&A...594A..31C}.
The high luminosity mode is characterised by:
a) a power law non-thermal component (accounting for the propeller shock);
b) a multi-temperature accretion disc, characterised by a free temperature and an inner radius (\texttt{diskbb} model). The disc temperature $T(r)$ is proportional to $r^{-p}$ (where $r$ is the disc radius and $p$ is a free parameter) and  we fixed $p$ = 0.5 to approximate an advection-dominated flow \citep{1994ApJ...428L..13N};
c) a heated polar cap with free radius and temperature (\texttt{nsatmos} model), to account for matter accreting onto the NS\footnote{ Technically, \texttt{nsatmos} model is for a cooling NS atmosphere, but is has been shown to describe fairly well also low-level accretion.}.

In the low(-soft) luminosity mode the radio pulsar emission and the disc component co-exist, so the model is more complex and is characterised by:
a) a power law due to a non-thermal emission ($\Gamma$), as in the high mode but with a different spectral index. This component might arise from the shock interaction between the incoming matter from the accretion disc and the pulsar wind;
b) a disc component, with the same $T(r) \propto r^{-0.5}$ curve, where the inner radius of the disc have to result larger than $r_{\rm lc}$;
c) a heated polar cap with the polar cap radius of the same size as in the high mode and allowing for the possibility of cooling; 
d) a secondary power ($\Gamma_P$) law accounting for the non-thermal magnetospheric pulsar emission.

The low-soft mode is considered here, because it occurs more frequently than the low-hard, and because it is more similar to the low mode of J1023.

Finally, the quiescent state is determined only by the pulsar emission (both thermal and non-thermal). 
All the parameters in the quiescent state (power law and polar cap radius including their normalisation) are then fixed to those of the pulsar in the low-soft mode. 
All spectra have been corrected for interstellar absorption using the same hydrogen column density (N$_H$).

\subsection{Pulsations from an intra-binary shock in a shock emission scenario}

The need for an alternative scenario comes from the discovery of optical pulsations in J1023 \citep{2017NatAs...1..854A}. 
Actually, there is no known mechanisms able to account for the large pulsed optical luminosity in J1023, involving the accretion of matter onto the NS surface (the best model, based on cyclotron emission falls by a factor $\gsim 35$ \citealt{2017NatAs...1..854A}).
\cite{2019arXiv190410433P} proposed that optical and X-ray pulsations are produced in intra-binary shocks when the relativistic wind of the pulsar interacts with the in-flowing matter close to the light cylinder. 
In the new physical picture (presented for J1023 also in \citealt{Campana2019}) the radio pulsar is always active, even during the X-ray active state, and the source of power for the pulsations is its rotational energy. The relativistic pulsar wind, interacting with the surrounding medium, produces high energy photons. Optical and X-ray pulsations are produced by synchrotron emission in the intra-binary shock, just beyond the light cylinder, at a distance $k\,r_{\rm lc}$, with $k$ =1-2 \citep{2019arXiv190410433P}. \cite{2019arXiv190602519V} also suggested that during the high luminosity mode the disc is truncated outside the light cylinder radius and the wind-disc interaction generates synchrotron radiation. The position of the termination shock can change. In particular, we expect that in the low luminosity mode the shock's radius expands, causing the disappearance of X-ray and optical pulsations. This hypothesis in J1023 is supported by the presence of radio flares in correspondence with the X-ray low luminosity modes \citep{2018ApJ...856...54B}. Radio flares are interpreted as rapid ejections of plasma by the active radio pulsar. \cite{2019arXiv190602519V} suggested instead that during the low  mode, matter is accreted onto the NS.
\cite{2019arXiv190410433P} suggested that flares might come from a complete enshrouding of the pulsar. At variance with the previous scenario, it seems natural here to include flare mode data into the spectral fits, to investigate this possibility.

The spectral model for this scenario could be parametrised by the following composite spectral model.
The pulsar is always active, also during the X-ray active state. For this reason all modes (flare, high, and low) and the quiescence are characterised by the NS thermal and non-thermal emission. The NS magnetospheric non-thermal emission is modelled by a power law with spectral index $\Gamma_{\rm mag}$. The NS thermal emission, produced by polar caps with free radius and temperature, is parametrised by the \texttt{nsatmos} model. All free parameters (spectral index $\Gamma_{\rm mag}$, NS temperature and radius) are equal for all modes and the quiescence.
In addition flare, high, and low modes are modelled by a second power law due to a further non-thermal emission, probably coming from the shock. The associated spectral index $\Gamma_{\rm shock}$ should be different in all modes because it depends on the geometry and density of the shock, hence it is a free parameter. 
A model with only these components is not able to reproduce the X-ray spectral data of J1023 (see \citealt{Campana2019}). 
One possibility is to include a partially ionised material enshrouding the pulsar modelled by the XSPEC model \texttt{zxipcf}. This should account for the presence of matter close to the pulsar. 
This spectral component depends on the amount of matter along the line of sight (equal for all three modes), the ionisation parameter $\xi~=L/nr^2$ (where $L$ is the source luminosity, $n$ the density of the medium and $r$ the distance of the medium), and the covering fraction $f$. Both $\xi$ and $f$ should be different for all three modes, varying luminosity, and distance of the shock, at least. 
The overall model is then composed as follows:
a) a power law non-thermal component accounting for the shock emission, different for the three modes (flare, high, and low modes);
b) a heated polar cap with free radius and temperature (\texttt{nsatmos} model) common to all modes and including the quiescent state, accounting for the pulsar polar cap thermal emission;
c) a power law non-thermal component common to all modes and including the quiescent state, accounting for the pulsar magnetospheric emission;
d) a partially ionised absorption component, different for the three modes, accounting for the inner edge disc absorption. 


\subsection{Low-soft and low-hard modes}
J12270, besides showing the distinctive behaviour between low and high modes characteristic of transitional pulsars, has also presented an unusual activity in the low mode, showing a soft and hard instance \citep{2010A&A...515A..25D, 2013A&A...550A..89D}.
Concerning the spectral analysis of the low-hard and low-soft modes, given the much lower statistics, we were forced to considered simpler spectral models.
Given that the low-hard mode is definitely brighter than the low-soft mode, we explored the possibility that the low-hard mode can be described as the superposition of an additional spectral component on top of the stable low-soft one.
We tested three different composite models. In all of these models we expect a common emission, described by a power law, and  an additional component for the low-hard spectra only, that causes the variation in spectral shape. 

We consider three different additive models to single power law describing the low-soft spectra. 
We add either a bremsstrahlung component (\texttt{bremss} or \texttt{mekal}, as for the tail of a magnetic flare),
a blackbody thermal emission (as footprints of hotter regions in the disc), or another power law (underlining our ignorance in the process). 

\section{Observations}
J12270 was observed four times by \textit{XMM-Newton} between 2009 and 2014. In our analysis, we considered only the first three observations (see Table \ref{table:2}) performed with the EPIC cameras (MOS and pn). We excluded the last observation (Obs. ID. 0729560801) since the source had a flux 1.6 times higher than the previous observation during quiescence \citep{2015MNRAS.454.2190D}. A variation in the quiescent flux of a MSP is not expected, nor observed in other MSPs and J1023. Given the construction of our models, we can only attain one single quiescent value. Fitting a variable quiescent flux would require the addition of another spectral component, which is not allowed by the quality of the data. We decided to show the results concerning the lowest quiescent flux. We repeated the same fitting procedure considering the brighter quiescent observation, finding that spectral parameters change by less than $10\%$ (see Appendix A). A follow-up discussion is presented in the last section.
In the first and third observation (Obs. ID. 0551430401 and 0727961401) both the EPIC-pn and the EPIC-MOS cameras operated in imaging full window mode using thin filters. The observation performed on 2011 Jan. 01 (Obs. ID. 0656780901) was made in fast timing mode with the EPIC-pn camera only.
The first two sets are widely discussed in \citet{2010A&A...515A..25D} and in \citet{2013A&A...550A..89D}. The last observation (2012 December 29) is described in \citet{2014ApJ...789...40B}, \citet{2015MNRAS.454.2190D}, and \citet{2015MNRAS.449L..26P}.

\begin{table*}
\centering
\caption{Summary of \textit{XMM-Newton} observations used in this analysis. MOS counts are the sum of MOS1 and MOS2 counts.}              
\label{table:2}      
\begin{tabular}{c c c c c c c c}          
\hline\hline                        
\tiny
Obs. ID. & Obs. start & Instrument & Duration$^{(*)}$ &Flare mode&  High mode&Low-soft mode & Low-hard mode\\
 & (YYYY-MM-DD) & & (ks)& dur.-ks (cts) & dur.-ks (cts) &  dur.-ks (cts) & dur.-ks (cts)\\   
\hline                                   
0551430401 & 2009-01-05 & pn(FF) &24.9&0.8 (6019) &16.7 (29662) & 2.7 (792) & 1.1 (694)\\      
 &  & MOS(FF) & 30.4& 0.9 (3408) & 18.6 (19155) & 3.0 (461) & 1.2 (455)\\
 \hline
0656780901 & 2011-01-01 & pn(FT)  &29.2& 1.1 (10778) &19.1 (71128) & 4.2 (3032) & 0.9 (1381)\\
 &  & MOS(FF)        & 30.6&1.1 (3486) &18.9 (20733) &4.1 (664) & 0.8 (426)\\
 \hline
0727961401 & 2013-12-29 & pn(FF)      & 27.9 & \multicolumn{3}{c}{Quiescence (1544)} \\
 & & MOS(FF) & 34.4& \multicolumn{3}{c}{Quiescence (1382)} \\
\hline
0729560801 & 2014-06-27 & MOS(FF) & 40.8& \multicolumn{3}{c}{Quiescence (2448)}\\
\hline

\end{tabular}

$^{(*)}$ The sum of exposures in different modes is lower than the total exposure time, since we selected pure modes, excluding the transition intervals.
\end{table*}

\section{Data reduction}
The data were processed with the $\textit{XMM-Newton}$ Science Analysis Software (SAS) version xmmsas$\_20170719\_1539$-$16.1.0$ and the latest calibration files. 
EPIC data were processed with $\texttt{emproc}$ and $\texttt{epproc}$ tasks. The data were filtered selecting pattern 0-12 (0-4) for MOS (pn) and FLAG == 0. 
We filtered the last observation (Obs. ID. 0727961401) that is affected by proton flares 
by excluding time intervals during which the overall count rate was greater than 10 cts s$^{-1}$ for the EPIC-MOS data, and greater than 50 cts s$^{-1}$ for the EPIC-pn data.
We used a region centred on the source with radius of 570, 676, and 312 pixel to extract EPIC-MOS (respectively for Obs. ID. 0551430401, 0656780901 and 0727961401) light curves and spectra. Background events were extracted from a similar region located on the same CCD with no sources.
In the first two observations, the source was too bright and EPIC-MOS data were affected by pile-up. To reduce this effect, we excised the innermost region of the source with a radius of 150 and 140 pixel (respectively for Obs. ID. 0551430401 and 0656780901).
The EPIC-pn events in fast timing mode were extracted from RAWX=29-50 px (Obs. ID. 0656780901) while events in imaging full window mode were extracted from 570 and 263 pixel region (Obs. ID. 0551430401 and 0727961401). No pile-up is present in the EPIC-pn data. 

\begin{figure}
  \resizebox{\hsize}{!}{\includegraphics{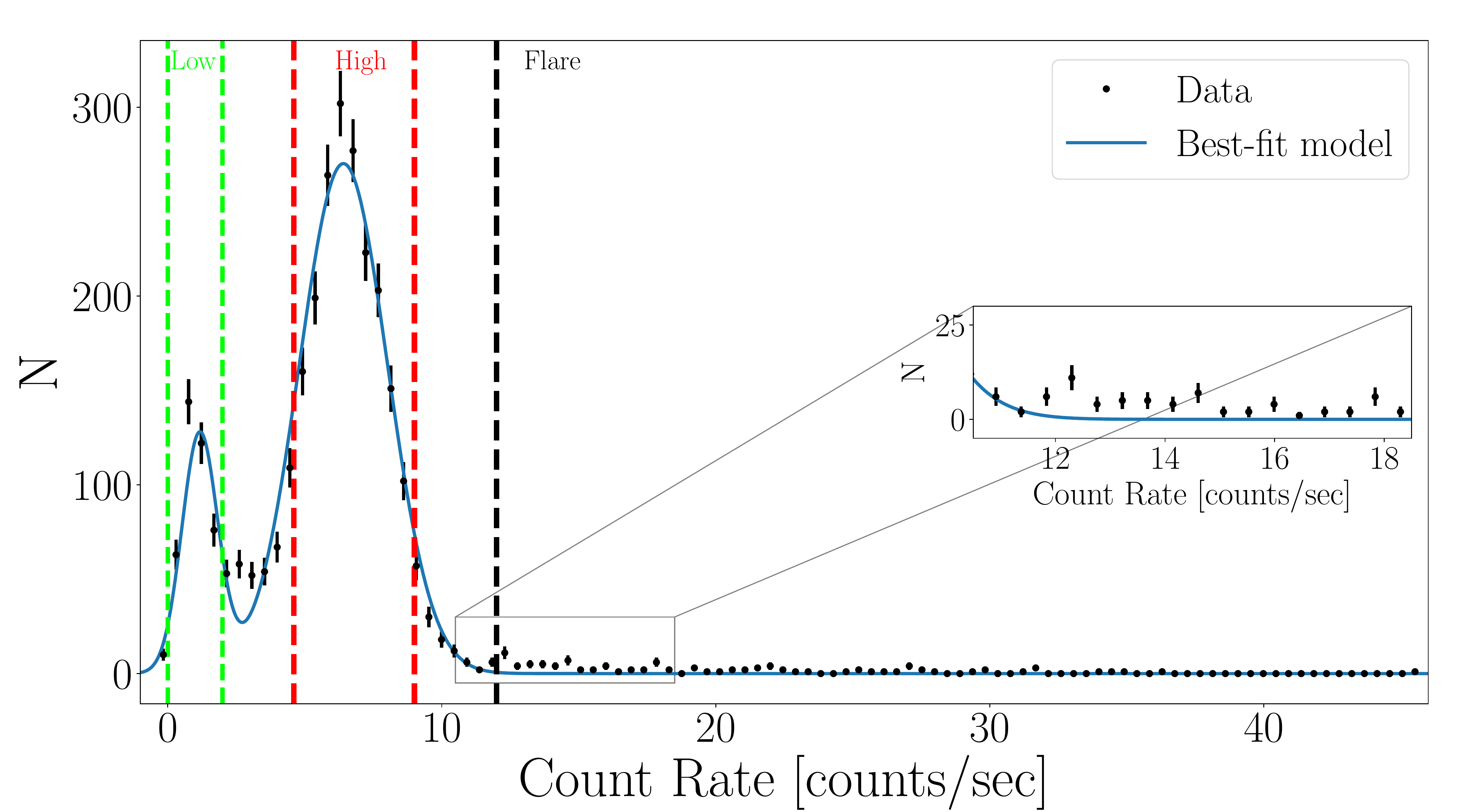}}
  \caption{Distribution of the background-subtracted J12270 count rate obtained from 20-s binned lightcurves from the 2009 January and 2011 January \textit{XMM-Newton} EPIC data. We define the region between 0 cts s$^{-1}$ and 2 cts s$^{-1}$ as the low-soft mode, while the region between 4.6 cts s$^{-1}$ and 9 cts s$^{-1}$ is the high mode. Higher than 12 count s$^{-1}$ we define flare mode. In the zoomed plot we can see that the double Gaussian fit does not interpolate data during the flares.}
  \label{histogram}
\end{figure}

\begin{figure}
  \resizebox{\hsize}{!}{\includegraphics{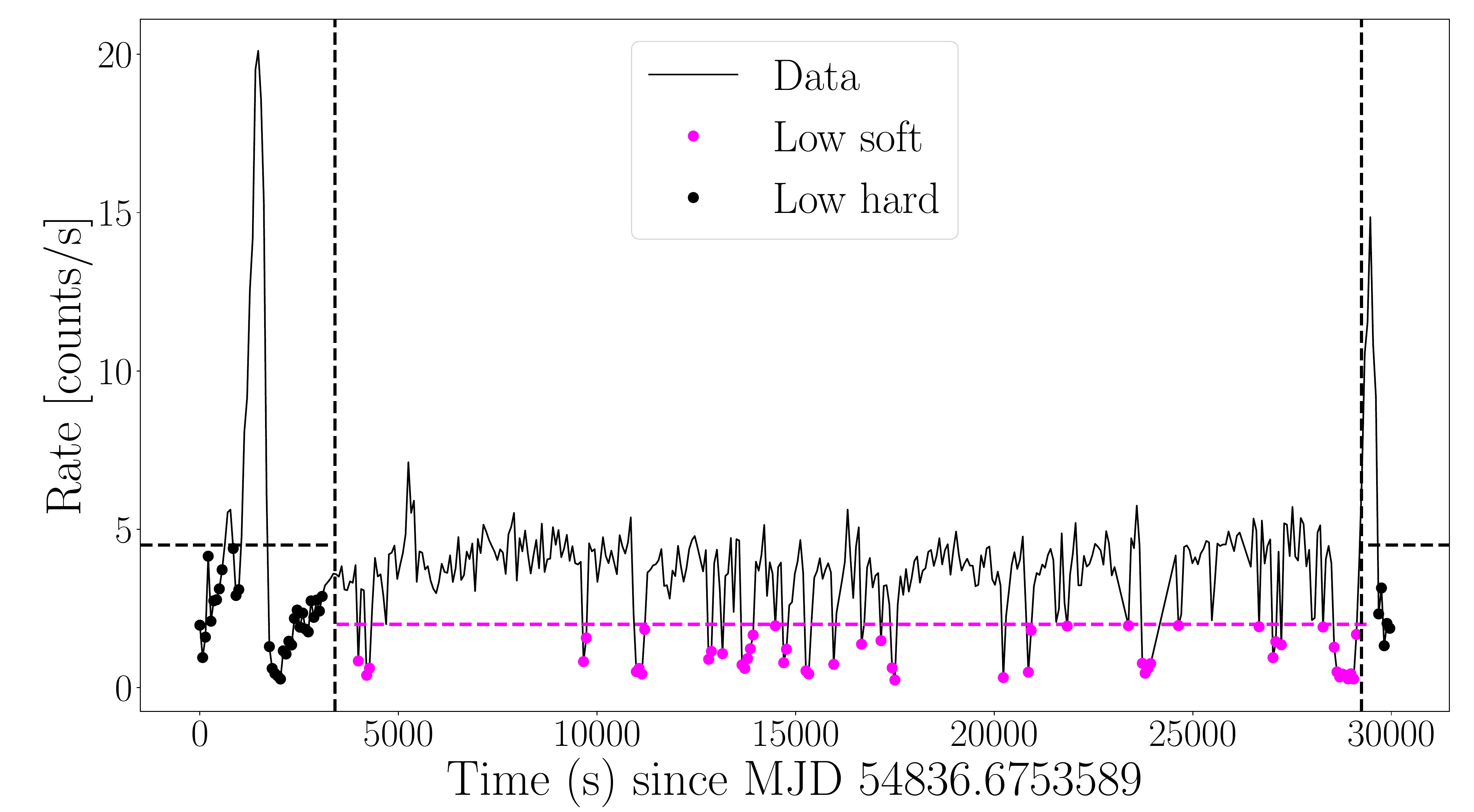}}
  \caption{EPIC-pn light curve (Obs. ID. 0551430401) in the 0.3-10 keV band with a time binning of 70 s. Magenta points refer to low-soft mode, while black points refer to low-hard mode.}
  \label{LC_oss1}
\end{figure}

\begin{figure}
  \resizebox{\hsize}{!}{\includegraphics{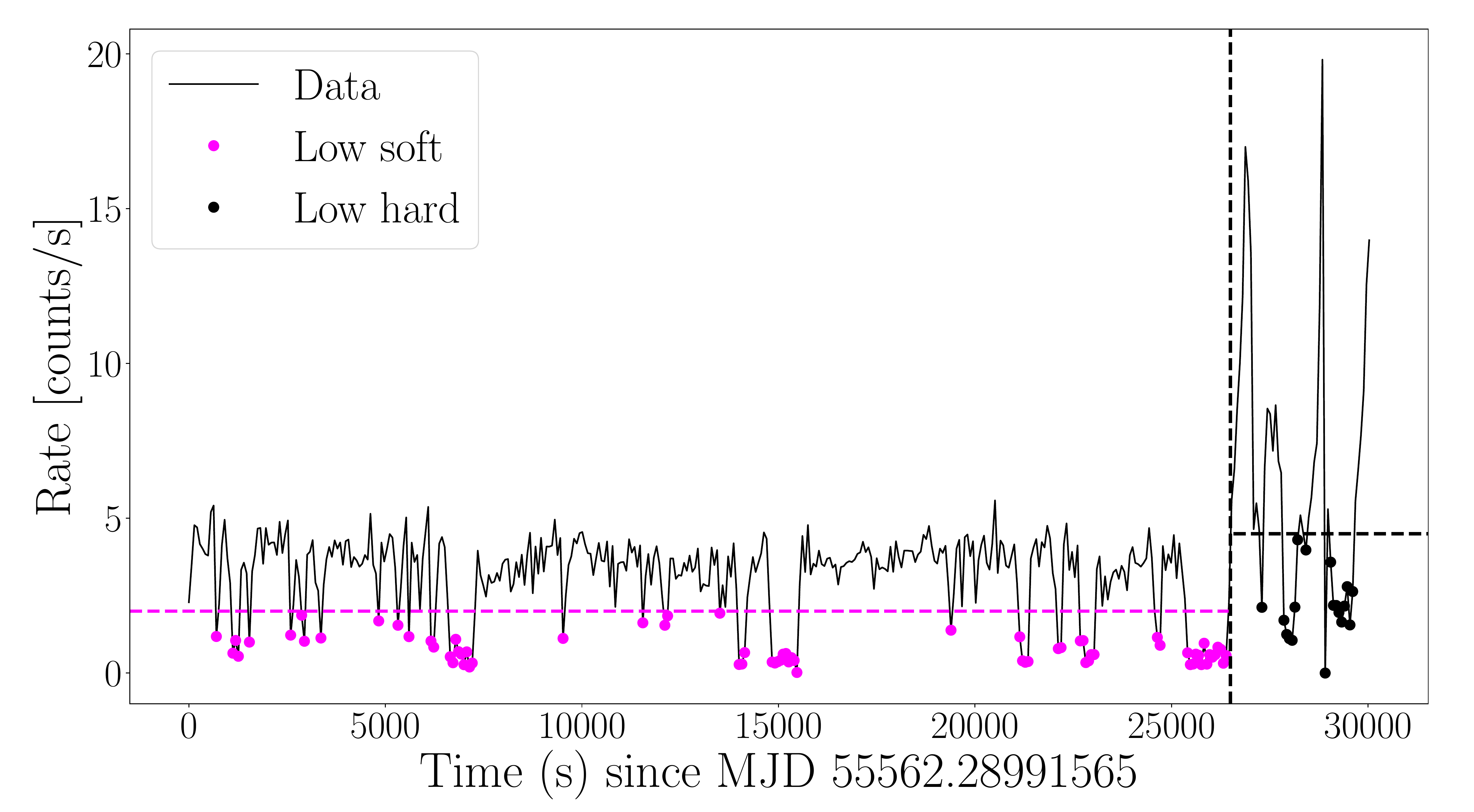}}
  \caption{EPIC-pn light curve (Obs. ID. 0656780901) in the 0.3-10 keV band with a time binning of 70 s. Magenta points refer to low-soft mode, while black points refer to low-hard mode.}
  \label{LC_oss2}
\end{figure}

For each observation we extracted source and background light curves of the three CCD cameras and we binned all of them with the same time resolution of 20 s. We combined background-subtracted and exposure-corrected EPIC 0.3-10 keV light curves by using the \texttt{epiclccorr} tool. 
The final light curves (see \citealt{2010A&A...515A..25D}, \citealt{2013A&A...550A..89D} and also Fig. \ref{LC_oss1}, Fig. \ref{LC_oss2}) are characterised by the three source modes described above. Fig. \ref{histogram} shows the distribution of count rates of the 2009 January and 2011 January observations. We defined the interval  0.0-2.0 cts s$^{-1}$ (selected in the time intervals 54836.70128259-54837.01239370 MJD for Obs. ID. 0551430401 and 55562.29653722-55562.61031037 MJD for Obs. ID. 0656780901) as the low-soft mode (see Fig. \ref{LC_oss1} and Fig. \ref{LC_oss2}), while the interval 4.6-9.0 cts s$^{-1}$ the high mode range.
We found a bimodal distribution of the count rates (as presented in \citealt{2015ApJ...806..148B} for J1023), with the low-soft and high mode seen as peaks at $\approx$ 1 cts s$^{-1}$ and $\approx$ 7 cts s$^{-1}$, respectively. 
For the low-soft and low-hard spectral analysis we defined the region between 0.0-4.5~cts~s$^{-1}$ as the low-hard mode. As seen in Fig. \ref{LC_oss1} and Fig. \ref{LC_oss2} low-hard modes are observed following flares and are selected only during the time intervals 54836.67414139-54836.70128259 MJD and 54837.01239370-54837.02171083 MJD for Obs. ID. 0551430401, 55562.61031037-55562.64217380 MJD for Obs. ID. 0656780901.

Spectra were extracted using intervals indicated above only for the first two observations because in the third observation the source was in a rotation-powered quiescent state. Redistribution matrix files (RMF) and ancillary response files (ARF) were extracted using \texttt{rmfgen} and \texttt{arfgen} tasks for all the spectra. ARF and RMF files for MOS1 and MOS2 data were summed for each observation. EPIC spectra were rebinned with FTOOL \texttt{grppha} in order to have 30 counts/bin. EPIC-MOS data were fitted in the 0.3-10 keV energy range, while (timing) EPIC-pn data in the 0.6-10 keV range.

\section{Spectral fitting}

We used XSPEC version 12.9.1p \citep{1996ASPC..101...17A} for fitting our data. In order to give a possible physical interpretation of state changes, we separated \textit{XMM-Newton} data into a flare mode (2 ks), a high mode (37 ks), a pure low mode (low-soft, 7 ks), and quiescent state (34 ks).

We corrected all the spectra for the absorption using \texttt{tbabs} model \citep{2000ApJ...542..914W} with \texttt{vern} cross sections \citep{1996ApJ...465..487V} and \texttt{wilm} abundances. We used the same absorbing column density for all the observations in the propeller-ejection model and in the shock emission model. 

\subsection{Propeller-ejection scenario}

\begin{figure}
  \resizebox{\hsize}{!}{\includegraphics{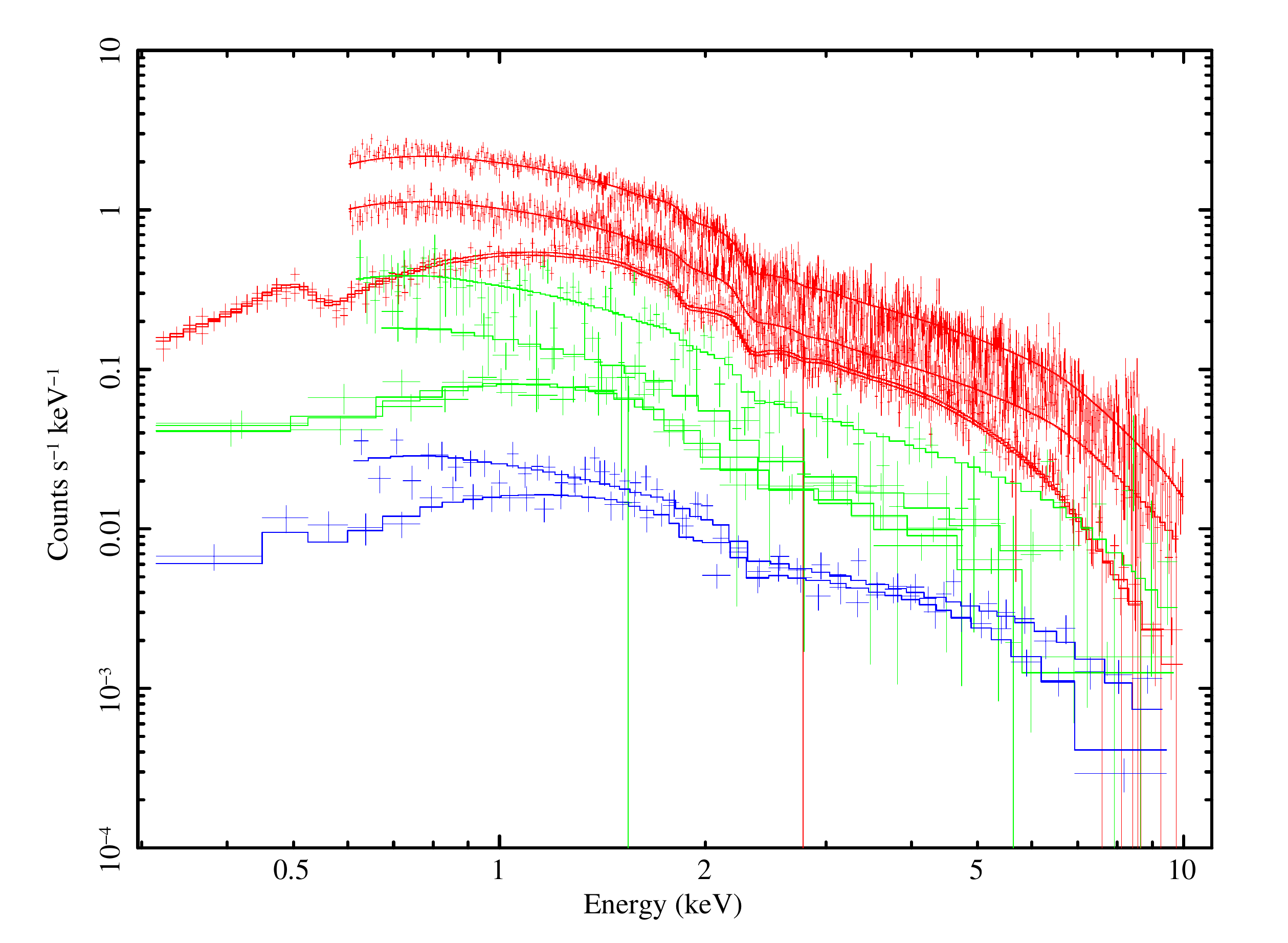}}
  \caption{\textit{XMM-Newton} EPIC-pn and EPIC-MOS spectra. Spectral fit of the propeller-ejection model. Red spectra are referred to the high mode, green spectra to the low mode, and blue one to the quiescent state. Spectra in energy range 0.6-10 keV are referred to the pn camera and spectra in energy range 0.3-10 keV are referred to the MOS camera.}
  \label{pulsar}
\end{figure}

In the spectral fit of high and low modes (see Fig. \ref{pulsar}), mainly due to poorer statistics as compared to J1023, 
we have to exclude the disc component since it is too weak and is not required to model the overall spectrum.  
We only estimate a limit for the disc temperature and the inner disc radius (see Table \ref{table:3}).  This has been estimated by adding this spectral component to the fit and derive the corresponding upper limits.
The \textit{XMM-Newton} spectral fit with the composite model described in Section 2.1 (excluding the accretion disc) worked well with $\chi^2_{red}=1.09$ for 2276 degrees of freedom\footnote{The statistics of the spectra is very high. In this situation, calibration uncertainties play a role. It is common practice to add (in quadrature) a $2\%$ systematic uncertainty to each spectral channel, as recommended in Smith et al. (2015) (see \url{http://xmm2.esac.esa.int/docs/documents/CAL-TN-0018.pdf}).} (see Table \ref{table:3} and Fig. \ref{pulsar}), resulting in a null hypothesis probability of 0.1\%. 
During the propeller state of J12770 (i.e. the high mode) the emission is dominated by a non-thermal component ($\sim$97\%) described by a power law with index $\Gamma=1.46 \pm 0.04$ (error calculated with $\Delta \chi^2$=2.71). A very small fraction of the flux ($\sim$3\%) is given by the NS thermal emission. In the radio-ejection state (i.e. the low-soft mode) the radio pulsar turned on and its flux contribution, due to non-thermal emission, achieved $\sim$18\% of the total unabsorbed flux. This contribution is described by a power law with index $\Gamma_P=1.0 \pm 0.1$. The dominant flux contribution ($\sim$81\%) also in this state is described by a power law with index  $\Gamma=1.67 \pm0.07$. A very low contribution ($\sim$1\%) is given by the NS thermal emission. 
In the radio pulsar state (i.e. the quiescence) the non-thermal emission is described by the same (by definition of the model) power law with $\Gamma_P$. 
This emission could come from the radio pulsar magnetosphere or from shocks, since we observed a modulation of X-ray emission at the orbital period. The NS thermal component contributed to 5.4\% of the total flux. 
{One might wonder if all these spectral components are needed. Indeed, they are all needed as far as are nested as specified in Section 2.1. The weakest component present in all the spectra is the thermal emission from the NS. Its significance is estimated to $4.4$-$\sigma$, by means of an F-test.}

\begin{table*}
\centering
\caption{Spectral fit parameters of  J12270 using the first composite model presented in Section 2.1. The fit provides a $\chi^2_{\rm red}$= 1.09 for 2276 degrees of freedom with an additional systematic error of 2$\%$. For each parameters errors are 90$\%$ confidence level ($\Delta \chi^2$ =2.71). NS emission radius is computed assuming a 1.4 kpc source distance, a 1.4 M$_{\odot}$ NS mass and a 10 km NS radius.}              
\label{table:3}      
\begin{tabular}{c c|c|c}          
\hline\hline                        
Parameter &High mode & Low mode & Quiescence\\   
\hline                                   
N$_H \ (10^{20}$ cm$^{-2})$ &9.5$^{+1.2}_{-1.1}$ & tied all & tied all \\
Power law $\Gamma$ &1.46$\pm$0.04 &1.67 $\pm$ 0.07 & -\\
Disc $T$ $^{(*)}$ (eV) &<100 &<89 & -\\
Disc norm N$_d$ $^{(*)}$& >220 & >339 &-\\
NS atmos. $T$ (eV) & 193$^{+42}_{-73}$ & 97$^{+22}_{-28}$ & tied  Low mode\\
NS atmos. Radius (km) & 1.0$^{+6}_{-1}$& tied all& tied all \\
NS power law $\Gamma_P$ & - &1.0 $\pm$0.1 & tied Low mode \\
\hline
\end{tabular}\\
$^{(*)}$ We estimated an upper limit for the disc components excluding them from the fit since the inefficient disc component is too weak for being constrained.
\end{table*}

\begin{figure}
  \resizebox{\hsize}{!}{\includegraphics{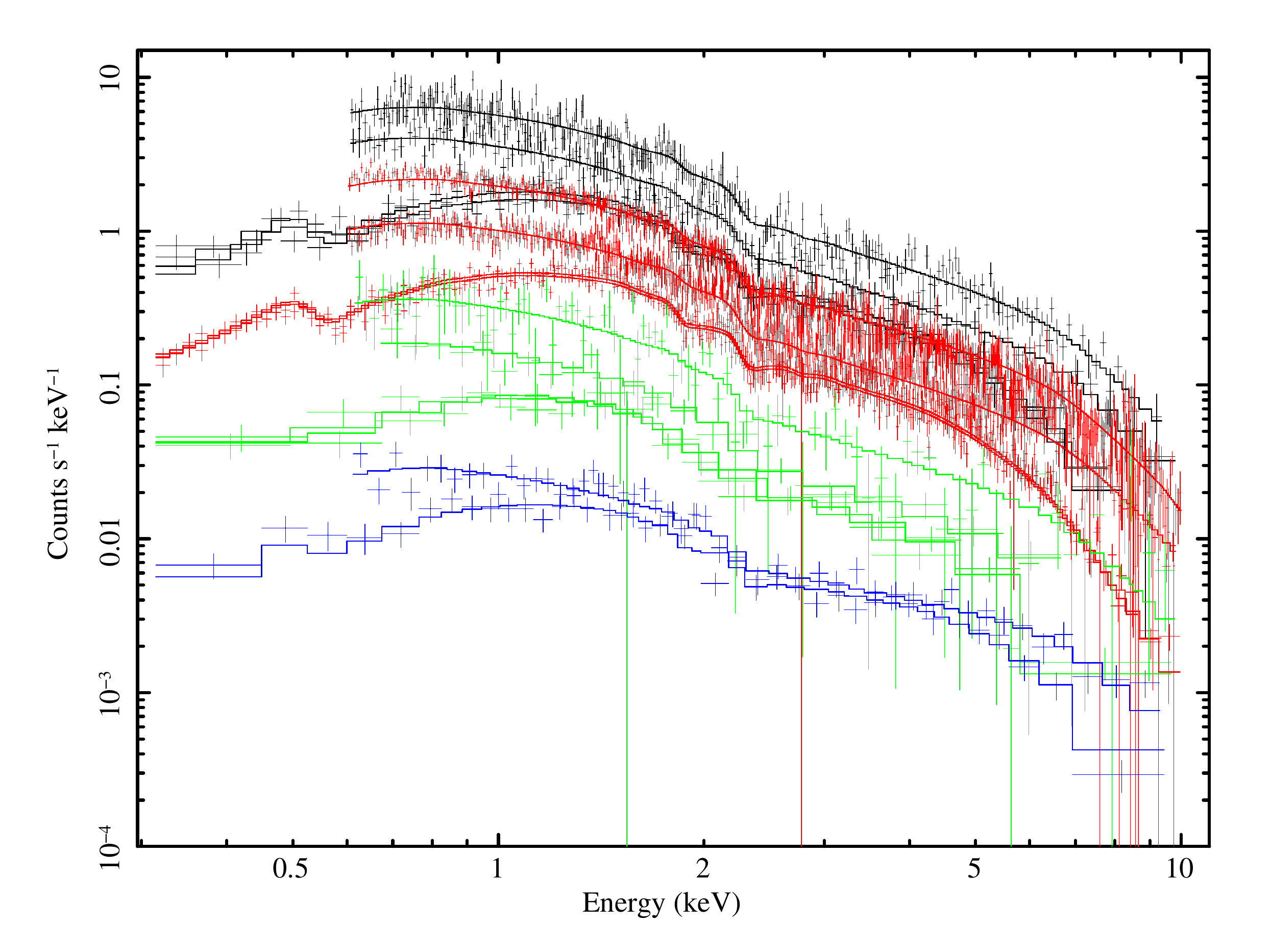}}
  \caption{\textit{XMM-Newton} EPIC-MOS spectra. Spectral fit of the shock emission model. Black spectra are referred to the flare mode, red spectra to the high mode, green spectra to the low mode, and blue one to the quiescent state. Spectra in energy range 0.6-10 keV are referred to the pn camera and spectra in energy range 0.3-10 keV are referred to the MOS camera.}
  \label{new_model}
\end{figure}

\subsection{Shock emission scenario}
In the shock emission scenario, as discussed in Section 2.2, we considered all modes (i.e. flare, high, low modes and quiescence, see Fig.~\ref{new_model}).
Following \cite{Campana2019} for J1023, we first selected only MOS data because the low energy part of the X-ray spectrum is very important to model the hot
absorption component (modelled within XSPEC with \texttt{zxipcf}). However, the statistics of the J12770 spectra is lower and the Galactic absorption higher with respect to the J1023 dataset to constrain this hot component. At variance with J1023, lacking here the soft absorption component, we included  in the spectral fits also the pn data. Therefore, we used exactly the same data for both scenarios for quiescence, low, and high mode data. 
Figure~\ref{new_model} and Table \ref{table:4} show the result of the composite spectral model described in Section 2.2 using both pn and MOS data. The spectral fit has a $\chi^2=1.07$ for 2877 degrees of freedom, with a null hypothesis probability of 0.2$\%$ (including a $2\%$ systematic uncertainty).
The NS emission, even if present in all states, appears significant only in the low mode and during the quiescence. In the low luminosity mode the magnetospheric emission achieved the 16.9$\%$ of the total unabsorbed flux, while during the quiescence the 93.5$\%$ (see Table \ref{table:5}). This non-thermal component is parametrised by a power law with a hard spectral index $\Gamma_{\rm mag}=1.0\pm0.1$. 
By construction, being the pulsar always active, the unabsorbed flux coming from the magnetospheric component is constant in high mode, low mode and quiescence ($\sim$0.9 $\times10^{-12}$ erg s$^{-1}$ cm$^{-2}$). Instead, the NS thermal component is irrelevant in the flare and high modes. It gives a low contribution in the low mode ($\sim$1$\%$) and the 7$\%$ of the total unabsorbed flux during the quiescence. The NS emission region has a temperature of 110$^{+45}_{-100}$ eV and the fraction of surface emitting is $f>0.01$.
In the flare and high luminosity modes the major flux contribution (99.3$\%$ and 97.2$\%$) is given by the second power law, associated with the shock emission. The relative spectral indices are $\Gamma_{\rm shock}=1.61\pm0.02$ for the flare mode and $\Gamma_{\rm shock}=1.54\pm0.01$ for the high mode. The high mode spectral index is consistent with the photon index $\Gamma=1.5$ predicted by a simple model of shock heating followed by fast synchrotron cooling of the post-shock plasma \citep{2019arXiv190602519V}. Also in this model the weakest component present in all the spectra is the thermal emission from the NS. Its significance is $3.3$-$\sigma$, by means of an F-test.
 
The residuals of the fit 
show the presence of a weak emission line close to 6.5 keV, that could indicate the presence of K$\alpha$ emission line during the X-ray active state. This line is expected in an accretion disc. We added a Gaussian to fit this component, with a line width fixed to zero, consistent with the instrumental spectral resolution. The K$\alpha$ line equivalent widths are equal to 65 eV in the flare mode, 59 eV in the high mode, and 362 eV in the low mode. We allowed the line normalisation to take negative and positive values in order to assess the Fe K$\alpha$ significance via the $F$-test \citep{Protassov2002}. The spectral fit improved significantly with the additional Gaussian component ($\Delta\chi^{2}$=23 for 3 degrees of freedom). 
The presence of this component is then supported by the $F$-test whose $p-$value is 5.8$\times$10$^{-5}$ (equivalent to $3.9$-$\sigma$). A more detailed study on statistical significance would require Monte Carlo simulations as presented in \cite{2018MNRAS.473.5680K}. We note that for such a large $\Delta\chi^2$ the line is always highly significant.

Finally, we tested what happens if we leave free the absorption column density in the flare mode: the two values (flare and all the other modes) are fully consistent.

\begin{table*}
\centering
\caption{Spectral fit parameters of J12270 using the second composite model presented in Section 2.2. The fit provides a $\chi^2_{\rm red}=1.07$ for 2877 degrees of freedom  with an additional systematic error of 2$\%$. For each parameters errors are 90$\%$ confidence level ($\Delta \chi^2$ =2.71).}              
\label{table:4}  
\begin{tabular}{c c|c|c|c}          
\hline\hline                        
Parameter &Flare mode&High mode &Low mode & Quiescence\\   
\hline                                   
N$_H \ (10^{22}$ cm$^{-2})$ &0.103$\pm$ 0.005&tied all & tied all & tied all \\
Power law $\Gamma_{\rm shock}$ &1.61$\pm$0.02 &1.54 $\pm$0.01 & 1.68 $\pm$0.07 & -\\
Thermal component log($T/K$) & tied all & tied all & tied all & 6.10$^{+0.17}_{-0.38}$\\
Emitting fraction $f$ & tied all& tied all& tied all &0.008$^{+0.35}_{-0.006}$ \\
NS power law $\Gamma_{\rm mag}$ &tied all &tied all & tied all& 1.0$\pm$0.1 \\
\hline
\end{tabular}\\
\end{table*}

\begin{table*}
\centering
\caption{0.3-10 keV flux of the three different luminosity modes and quiescence of J12270. Percentage of the contribution of spectral components in the first and the second model.  Fluxes refer to the mean of the MOS detectors. Unabsorbed luminosities are calculated assuming a distance equal to 1.4 kpc.}              
\label{table:5}      
\begin{tabular}{c c c|c|c|c}          
\hline\hline                        
Parameter & Total/&Flare mode &High mode &Low mode & Quiescence\\
 & spectral components & & & & \\  
\hline
 & & \multicolumn{4}{c}{\textbf{Propeller-ejection model}}\\
\hline
Absorbed flux (10$^{-12}$ erg s$^{-1}$ cm$^{-2}$) & Total& - & $16.5^{+0.9}_{-0.9}$  &$2.4^{+0.2}_{-0.2}$ & $0.43^{+0.02}_{-0.03}$ \\
\hline
Percentage of contribution to& Power law & &97.2 $\%$ & 81.1 $\%$ & - \\
the unabsorbed flux & Ineffic. disc$^{(*)}$ & &0.0 $\%$& 0.0 $\%$& - \\
  & NS thermal & &2.8 $\%$ &1.1 $\%$& 5.4 $\%$\\
 & Magn. Power law & &- & 17.8 $\%$ & 94.6 $\%$\\
\hline 
Unabsorbed flux (10$^{-12}$ erg s$^{-1}$ cm$^{-2}$) & Total & - &18.1 & 2.7 & 0.46 \\
Luminosity (10$^{32}$ erg s$^{-1}$) & Total & - &42.2  & 6.2  & 1.1 \\
\hline
 & & \multicolumn{4}{c}{\textbf{Shock model}}\\
\hline
Absorbed flux (10$^{-12}$ erg s$^{-1}$ cm$^{-2}$) & Total& $36.3^{+1.1}_{-1.1}$ & $16.3^{+0.2}_{-0.2}$  &$2.4^{+0.1}_{-0.1}$ & $0.43^{+0.03}_{-0.04}$ \\
\hline
Percentage of contribution to& Shock emission & 99.3$\%$ &97.2$\%$ &81.8$\%$& -\\
the unabsorbed flux & NS thermal & 0.0$\%$ &0.2$\%$ &1.3$\%$&6.5$\%$\\
  & Magnetospheric emission & 0.7$\%$ & 2.6$\%$ &16.9$\%$ &93.5$\%$\\
\hline
Unabsorbed flux (10$^{-12}$ erg s$^{-1}$ cm$^{-2}$) & Total & 40.9 &18.1 & 2.7 & 0.46 \\
Luminosity (10$^{32}$ erg s$^{-1}$) & Total & 95.5 &42.4  & 6.4  & 1.1 \\
\hline                                            
\end{tabular}\\
$^{(*)}$ We obtained only an upper limit on the 0.3-10 keV flux of the thermal inefficient disc component.\\
Fluxes refer MOS data.
\end{table*}

\begin{table*}
\centering
\caption{Spectral fit parameters of J12270 low mode spectra (see Section 5.3). Errors are 90$\%$ confidence level ($\Delta \chi^2$ =2.71) for each parameters. The fit provides a $\chi^2_{\rm red}=1.20$ with 237 degrees of freedom.}              
\label{table:6}      
\begin{tabular}{c c|c|c}          
\hline\hline                        
Parameter & Low-soft mode&Low-hard mode  &Low-hard mode \\
          &                   &  (Obs. ID. 0551430401) & (Obs. ID. 0656780901) \\ 
\hline                                   
N$_H \ (10^{22}$ cm$^{-2})$  & 0.11$\pm $0.03 & 0.8$_{-0.2}^{+0.3}$ & 0.20$_{-0.1}^{+0.1}$   \\
Power law $\Gamma_1$ &1.62 $\pm$ 0.09  & tied all  & tied all  \\
Power law $\Gamma_2$ & -- & 0.2$_{-0.2}^{+0.2}$ & 0.5$_{-0.2}^{+0.1}$ \\
\hline
\end{tabular}
\end{table*}

\subsection{Soft and hard low modes}

Concerning the analysis of low-soft and low-hard modes, we tried the three spectral models discussed in Section 2.3.
The first model with the additional bremsstrahlung emission (see Section 2.3) does not reproduce the data: for \texttt{mekal} model we got $\chi^2_{\rm red}=3.97$ for 238 degrees of freedom.
The addition of a black body or a second power law to the power law did not produce an acceptable fit either: $\chi^2_{\rm red}~=1.51$ or $\chi^2_{\rm red}~=1.76$ for 239 degrees of freedom, respectively.

We then left the column density of the low-hard spectra free to vary with respect to the low-soft spectra. The fit with the addition of a black body worked well, with a $\chi^2_{\rm red}~=1.05$ for 238 degrees of freedom and a null hypothesis probability of 27.7\%. However, the radius of the emitting region is very small ($\sim$ 1 km) and with a very high temperature ($\sim$ 10$^7$ K). The physical reason for such a small radius and high temperature is difficult to envisage.

\begin{figure}
  \resizebox{\hsize}{!}{\includegraphics{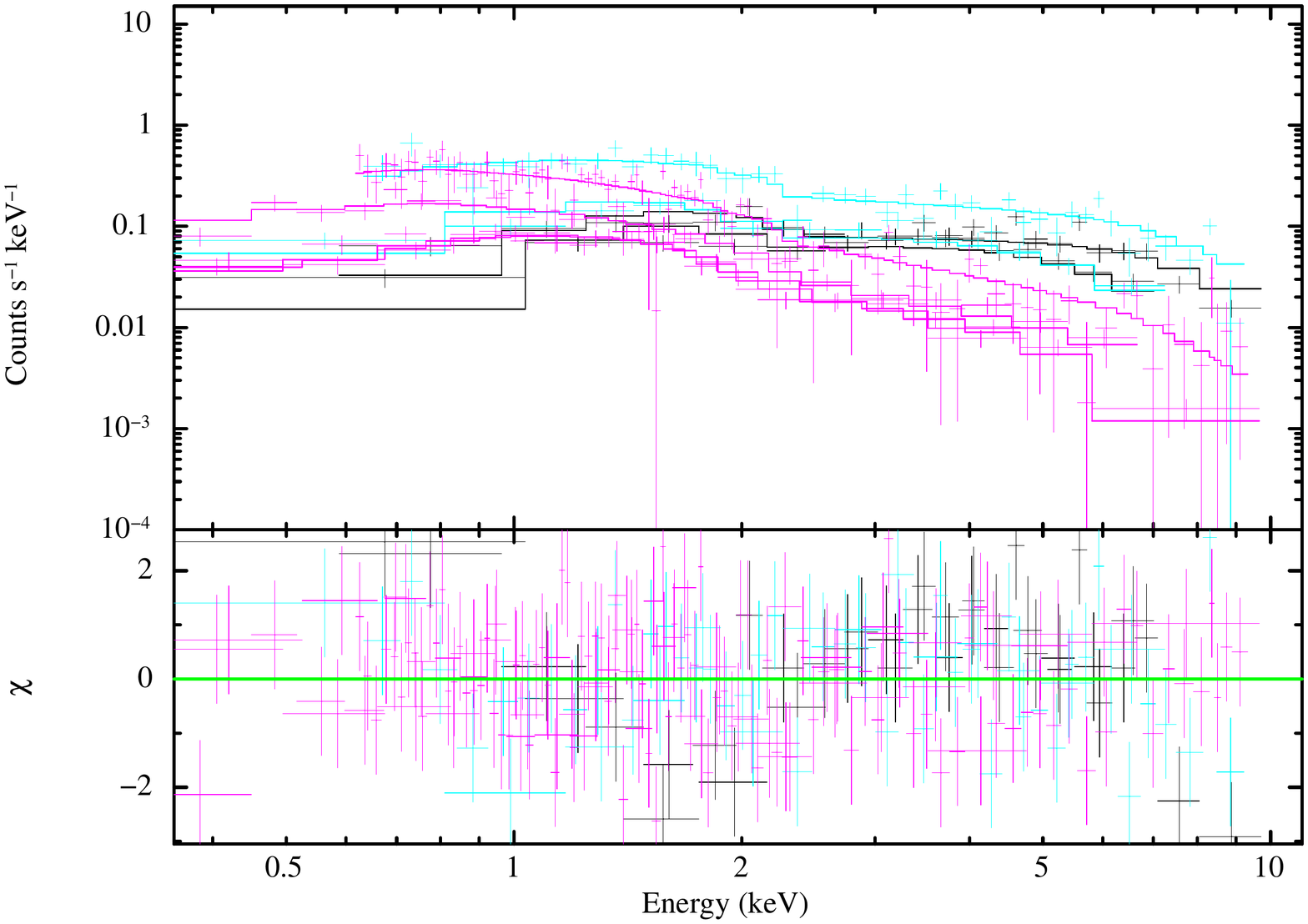}}
  \caption{\textit{XMM-Newton} EPIC-pn and EPIC-MOS spectra. Black spectra are referred to the low-hard mode (Obs. ID. 0551430401), light blue spectra to the low-hard mode (Obs. ID. 0656780901) and magenta one to the low-soft mode. Spectra in energy range 0.6-10 keV are referred to the pn camera and spectra in energy range 0.3-10 keV are referred to the MOS camera. The residuals are plotted in the low panel.}
  \label{hardsoft}
\end{figure}

The best composite model is the one with the sum of two power laws and free column density for the low-hard spectra. The spectral fit (see Fig. \ref{hardsoft} and Table \ref{table:6}) has a $\chi^2_{\rm red}=1.20$ for 237 degrees of freedom and a null hypothesis probability of 1.7\%. The absorption $N_H$ in the low-hard mode is consistent with the low-soft one for the second observation and much higher (factor of $\sim 7-10$) during the first observation. This is hard to interpret, too.
The first power law, equal for the two modes (low-hard and low-soft), has an index $\Gamma_1=1.62 \pm 0.09$, consistent with the value $\Gamma=1.67 \pm 0.07$ computed in the propeller-ejection scenario in the low mode (see Table \ref{table:3}) and with $\Gamma_{\rm shock}=1.68\pm0.07$ in the shock emission scenario (see Table \ref{table:4}). The second power law varies for the two low-hard spectra and in both cases is very hard with an index $\Gamma_2=0.2 \pm 0.2$ for the first observation (Obs. ID. 0551430401) and $\Gamma_2=0.5_{-0.2}^{+0.1}$ for the second observation (Obs. ID. 0656780901).

\section{Discussion and conclusions} 

We analysed the X-ray spectra in the 0.3-10 keV range of the transitional pulsar XSS J12270$-$4859. Our observations were taken before and after the state transition from X-ray active state to quiescence (radio pulsar) state, which took place in Nov-Dec 2012.

We proposed two different physical scenarios to explain the rapid variability between high and low modes.
In the first scenario, the high-low mode transitions 
are accounted for by propeller to radio-ejection states (with the X-ray pulsations accretion-powered), in the second scenario (shock emission scenario) by an approaching or receding inner disc just outside the light cylinder (with the X-ray pulsations rotation-powered).

These scenarios were developed for J1023 (\citealt{2016A&A...594A..31C}; \citealt{Campana2019}).
The two spectral models fit the overall set of {\it XMM-Newton} observations sufficiently well, especially considering the large amount of data.
There is a slightly small statistical preference for the shock emission scenario. At variance with J1023, we do not detect here the presence of an accretion disc in the propeller-radio ejection scenario, nor do we find signs of a hot absorber in the shock emission scenario. This is mainly due to the lower statistics with respect to J1023, which continues to be in the X-ray active state, whereas J12270 is now in the quiescent state.
Contrary to J1023, we find signs of the presence of an iron K$\alpha$ emission line close to 6.5 keV, in the shock emission scenario. This might originate in the innermost region of the accretion disc.

Another difference with J1023 is the presence of two different spectral instances of the low mode. 
We performed a spectral analysis by comparing the different spectral shapes during the low-soft and low-hard modes, in order to try to understand  the physical process that drives the spectral change. 
We considered a spectral model with one component common to the two modes and an additional one for the low-hard mode only. We took a power law as common component. 
Only the addition of a second power law spectrum for the low-hard mode (see Table \ref{table:6}) and a variable absorption column density gave acceptable spectral fits.
We think that this harder component can be correlated with the tail of the flare emission, even if the variable absorption is difficult to explain.

It is important to note that a transition to the low mode occurred also during a flare. This implies that the emission mechanism related to low mode is able to temporarily stop or at least perturb the flare, even if not completely, since some emission remains (i.e. low-hard mode). After this, the flare is able to restart again, so that the quenching induced by the low mode transition is not ultimate. At the same time a tail or a residual emission of the flare is still present during the low-hard mode, meaning that the quenching is not complete or instantaneous. This provides a clear indication that flares are unrelated to high-and-low mode transitions. 

\begin{figure}
  \resizebox{\hsize}{!}{\includegraphics{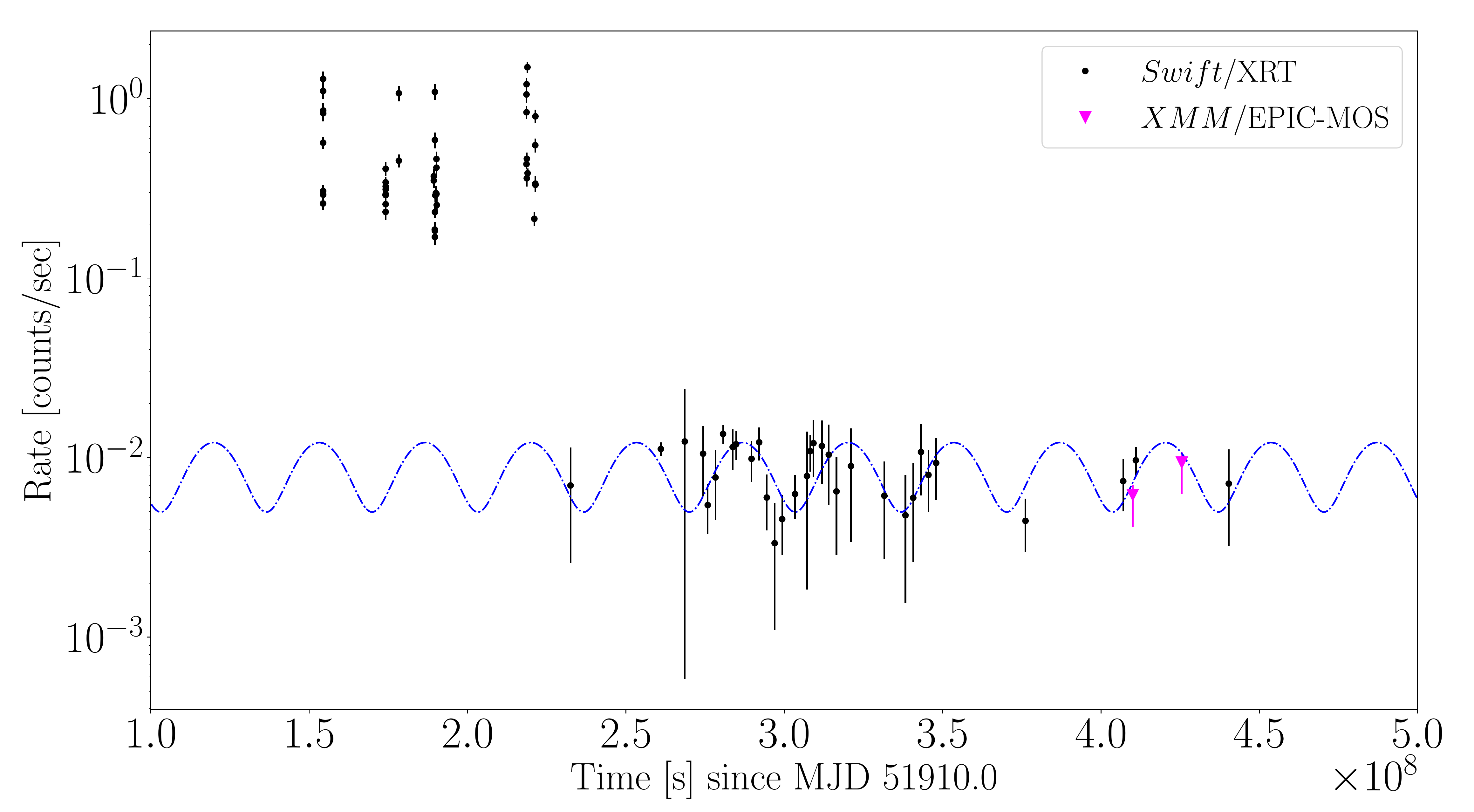}}
  \caption{\textit{Swift}/XRT count rate light curve of J12270. The two {\it XMM-Newton} observations were plotted with magenta triangles. These data were extrapolated to XRT counts through spectral modelling, including the uncertainty in the {\it XMM-Newton} flux. The overall X-ray light curve shows J12270 bright in the X-ray band at early times, and then a turn off to quiescence. During quiescence, a clear variability is apparent in the light curve.}
  \label{xrt}
\end{figure}

J12270 is currently in its quiescent state, with pulsations detected in the radio band outside extended eclipses close to the pulsar's superior conjunction. 
In our analysis we excluded the last \textit{XMM-Newton} observation (Obs. ID. 0729560801), performed in the radio pulsar state since it shows a slightly increased flux (factor $\sim 1.6$) with respect to a similar observation (Obs. ID. 0727961401) taken one year before during the radio pulsar state. 
We discarded this observation, because our primitive scenarios allows only for a fixed quiescent luminosity. 
We investigated in more details this puzzling behaviour. We built the {\it Swift} XRT light curve with the on-line generator\footnote{https://www.swift.ac.uk/user\_objects/} \citep{evans09}  (see Fig. \ref{xrt}). Interestingly, the quiescent light curve shows a variability by $\sim 40\%$ in the count rate. We also converted the two quiescent {\it XMM-Newton} fluxes into XRT count rates, including in the error budget the uncertainties in the flux, providing two more points.
The {\it XMM-Newton} count rate was converted into a {\it Swift}/XRT rate using PIMMs (v4.10b) and the above spectral parameters, and including the error on the {\it XMM-Newton} flux (conveying the spectral fit uncertainties), which provides the major contribution to the error budget.
A fit with a constant value provides an unsatisfactory fit with a reduced $\chi^2=1.8$ with 34 d.o.f. A sinusoidal modulation provides a $3.5\,\sigma$ improvement in the fit by means of an F-test ($\chi^2=1.1$ with 31 d.o.f.). The period is $387\pm7$ d. 
We note that the two {\it XMM-Newton} observations almost sample the entire rate variability. 
We can only speculate on the nature of this putative periodicity, whose interpretation is not straightforward.
One possibility is the precession of the neutron star with an asymmetric relativistic wind, which will then interact with different efficiencies with the material outflowing from the companion. 
Further observations and monitoring would be needed to better investigate this periodicity.

\section*{Acknowledgements}
We are deeply indebted with the referee for a careful reading of the manuscript and useful suggestions. AMZ and AR acknowledge the support of the PHAROS COST Action (CA16214). SC and PDA acknowledge support from ASI grant I/004/11/3. AR gratefully acknowledges financial support by the research grant ``iPeska'' (P.I. Andrea Possenti) funded under the INAF national call PRIN-SKA/CTA approved with the Presidential Decree 70/2016. AMZ thanks F. Coti Zelati, A. Papitto, and C. Righi for their help in data analysis.
AMZ would also like to thank G. Benevento for comments on an early draft and for discussions. This work made use of data supplied by the UK Swift Science Data Centre at the University of Leicester.

\begin{appendix}
\section{Spectral fittings with brighter quiescent observation}

\begin{table*}
\centering
\caption{Spectral fit parameters of J12270 using the first composite model presented in Section 2.1. The fit provides a $\chi^2_{\rm red}$= 1.1 for 2255 degrees of freedom with a a null hypothesis probability of 0.08 \% and an additional systematic error of 2$\%$. For each parameters errors are 90$\%$ confidence level ($\Delta \chi^2$ =2.71). NS emission radius is computed assuming a 1.4 kpc source distance, a 1.4 M$_{\odot}$ NS mass and a 10 km NS radius.}              
\label{table:A1}      
\begin{tabular}{c c|c|c}          
\hline\hline                        
Parameter &High mode &Low mode & Quiescence\\   
\hline                                   
N$_H \ (10^{20}$ cm$^{-2})$ &9.6$^{+1.2}_{-1.1}$ & tied all & tied all \\
Power law $\Gamma$ &1.47$^{+0.03}_{-0.04}$ &1.78$^{+0.28}_{-0.18}$ & -\\
Disc $T$ $^{(*)}$ (eV) &<100 &<89 & -\\
Disc norm N$_d$ $^{(*)}$& >220 & >339 &-\\
NS atmos. $T$ (eV) & 188$^{+43}_{-73}$ & 99$^{+33}_{-40}$ & tied Low mode\\
NS atmos. Radius (km) & 1.0$^{+7}_{-1}$& tied all& tied all \\
NS power law $\Gamma_P$ & - &1.0 $\pm$0.1 & tied Low mode \\
\hline
\end{tabular}
\leftline{$^{(*)}$ We estimated an upper limit for the disc components excluding them from the fit since the inefficient disc component is too weak}\\
\leftline{for being constrained.}
\end{table*}

\begin{table*}
\centering
\caption{Spectral fit parameters of J12270 using the second composite model presented in Section 2.2. The fit provides a $\chi^2_{\rm red}=1.07$ for 2856 degrees of freedom  with a a null hypothesis probability of 0.3 \% and an additional systematic error of 2$\%$. For each parameters errors are 90$\%$ confidence level ($\Delta \chi^2$ =2.71).}              
\label{table:A2}  
\begin{tabular}{c c|c|c|c}          
\hline\hline                        
Parameter &Flare mode& High mode &Low mode & Quiescence\\   
\hline                                   
N$_H \ (10^{22}$ cm$^{-2})$ &0.106$\pm$ 0.006&tied all & tied all & tied all \\
Power law $\Gamma_{\rm shock}$ &1.62$\pm$0.02 &1.57 $\pm$0.02 & 1.93 $^{+0.28}_{-0.20}$ & -\\
Thermal component log($T/K$) & tied all & tied all & tied all & 6.24$^{+0.15}_{-0.26}$\\
Emitting fraction $f$ & tied all& tied all& tied all &0.005$^{+0.02}_{-0.004}$ \\
NS power law $\Gamma_{\rm mag}$ &tied all &tied all & tied all& 0.9$\pm$0.1 \\
\hline
\end{tabular}\\
\end{table*}

Here we report the spectral fit parameters of J12270 using the brighter quiescent observation (Obs. ID. 0729560801) instead of the fainter one (Obs. ID. 0727961401). This observation performed on 2014 June 27 (see Table \ref{table:2}) was
made in fast timing mode with the EPIC-pn camera and in imaging full window mode using thin filters with the EPIC-MOS cameras. In these spectral fits we selected only MOS data because during quiescence the source is too faint to be detected with EPIC-pn camera operating in timing mode. 
In Table \ref{table:A1} we report the results of the first composite model presented in Section 2.1 and in Table \ref{table:A2} the results of the second composite model described in Section 2.2. Spectral parameters change by less than
10\% using the second observation during the quiescence.

\end{appendix}

\end{document}